# Ultrafast Polaron Dynamics in Layered and Perovskite Manganites: 2D and 3D Polarons

J. Lee,[1,*] H. J. Lee,[2] R. D. Averitt,[3] S. A. Trugman,[4] J. Demsar,[5] D. J. Funk,[6] N. H. Hur,[7] Y. Moritomo,[8] A. J. Taylor,[1] R. P. Prasankumar,[1] D. A. Yarotski[1,**]

[1]Department of Physics Education and Research Institute of Natural Science, Gyeongsang National University, Jinju-si, 52828, South Korea
[2]SLAC National Accelerator Laboratory, Menlo Park, CA 94025
[3]University of California, San Diego, CA 92024, USA
[4]Theoretical Division, Los Alamos National Laboratory, Los Alamos, NM 87545
[5]Institute of Physics, University of Mainz, Mainz, Germany
[6] Los Alamos National Laboratory, Los Alamos, NM 87545
[7]Center for CMR Materials, Korea Research Institute of Standards and Science, Daejeon, 305-600, Korea
[8]Department of Applied Physics, Nagoya University, Japan
[*] Email: lee.phys.edu@gnu.ac.kr
[**] Email: dzmitry@lanl.gov

We have studied the sub-picosecond quasiparticle dynamics in the perovskite manganite $La_{0.7}Ca_{0.3}MnO_3$ and the layered manganite $La_{1.4}Sr_{1.6}Mn_2O_7$ using ultrafast optical spectroscopy. We found that for $T \geq T_C$, initial relaxation proceeds on the time scale of several hundred femtoseconds and corresponds to the redressing of a photoexcited electron to its polaronic ground state. The temperature and dimensionality dependence of this polaron redressing time provides insight into the relationship between polaronic motion and spin dynamics on a sub-picosecond time scale. We also observe a crossover to a more conventional electron-phonon relaxation in the ferromagnetic metallic phase below $T_c$.

Although interest in the unusual electronic and magnetic properties of "Colossal" negative magnetoresistance (CMR) materials has led to a tremendous research effort both theoretically [1] and experimentally [2], there are still many unresolved issues related to the behavior of CMR materials. Basic mechanisms of the ferromagnetic metallic state emergence in manganites have been understood within the framework of double exchange theory [3] as shown in Fig. 1(a). In addition to double exchange, mechanisms such as the Jahn-Teller (JT) distortion with polaron formation (Fig. 1(b)), charge localization, and phase separation [4] have been proposed to interpret various phenomena in CMR materials. However, most of these mechanisms have been understood from a static, as opposed to dynamic perspective. For a complete understanding of the behavior of CMR materials, particularly near the transition temperature, it is essential to investigate their dynamic non-equilibrium properties.

Y. Moritomo *et al* [2] experimentally found that magneto resistivity (MR) of layered manganese oxide show larger magnetic fluctuations up to temperatures significantly above $T_c$ comparing with 3 D manganese. Motivated by this experiment, our main purpose of this paper is to characterize the dynamical properties of polarons in 2 and 3 dimensions.

The charge and lattice dynamics and interactions can be described by Holstein Hamiltonian which includes double exchange interaction and is expressed as [5]

$$H = H_{el} + H_{el-ph} + H_{ph}$$
$$= -\sum_{ij}(t_{ij}C_i^+ C_j + t_{ji}C_j^+ C_i) - \lambda \sum_j C_j^+ C_j (a_j + a_j^+) + \omega \sum_j a_j^+ a_j \quad (1)$$

Where $C_i^+$ and $a_i^+$ are the electron and phonon creation operators at site $i$. The effective hopping matrix element $t_{ij}$ between site $i$ and $j$ in the first term is determined by the angle between the local spins $\mathbf{S}_i$ and $\mathbf{S}_j$ as $t_{ij}=t\,cos(\theta_{ij}/2)$ through the double exchange integration [6]. In the Holstein model, the electrons are assumed to be coupled to intra-molecular deformation, which is a dispersionless optical phonon mode ($a_j^+ + a_j$), with the coupling constant $\lambda$ as shown in the second term in Eq. (1). The third term describes phonon energy with frequency $\omega$. The above Hamiltonian describes a polaron which consists of a localized charge with its associated lattice distortion field and as a results, can be easily detected by diffuse scattering around the Bragg peaks in neutron and x-ray scattering experiments [7,8,9].

Polaron, self-trapped by local lattice distortions, and its competition with magnetic and electronic ground states can explain the colossal magnetoresistance in transition metal oxides such as $La_{1-x}A_xMnO_3$ (A=Sr, Ca, Ba).[1] A proper hole doping in such system results in conducting ferromagnetic behavior below Curie temperature ($T_C$). The basic microscopic mechanism responsible for this behavior is caused by the double exchange interaction being stronger than electron-phonon interaction thus leading to electron self-trapping by distorted lattice when neighboring ionic spins are not parallel, as shown in Eq. 1 and Fig. 1(b). Figure 1(a) shows that the $e_g$ electron on a $Mn^{3+}$ site can hop to a neighboring $Mn^{4+}$ which has empty $e_g$ shell. However there is strong Hund's coupling between the $e_g$ electron and three $t_{2g}$ electrons, which want to keep them all aligned and prevents electron hopping if they are not. Double exchange interaction in high conductivity in ferromagnetic state wins over the electron-phonon coupling and leads high conductivity (the first term in Eq. (1)). However, at $T>T_C$, double-exchange term in Eq.1 decreases and electron-phonon interaction becomes prevalent and results in paramagnetic insulating state. In undoped manganese perovskites, such as $LuMnO_3$, each $Mn^{3+}$ has four d-electrons as shown in Fig. 1(b). Octahedron of $Mn^{3+}O_6$ spontaneously distorts (Fig. 2(b)) because an increased elastic energy is compensated by a resulting electronic energy savings due to the distortion. This is called Jahn-Teller effect and this electron-lattice coupling results in insulting properties of the system in addition to a suppressed double-exchange process at $T>T_C$.

Ultrafast optical spectroscopy is a powerful tool for investigating quasiparticle dynamics because various interactions can be temporally resolved at the fundamental timescales of nuclear and electronic motion [4, 10-13]. Previously, our group has identified the polaronic signatures in early dynamics of $La_{0.7}M_{0.3}MnO_3$(M= Ca, Sr) using optical pump-optical and terahertz (THz) probe spectroscopies [11]. These results show that THz conductivity increases right after strongly coupled electron-lattice state (polaron) is destroyed by optical excitation near the magnetic transition temperature, $T_C$ (Fig. 2(b) in ref. [11], where we did not focus on the dynamics of polaron itself). We have also studied dynamics of nanoscale phase separation in $Nd_{0.5}Sr_{0.5}MnO_3$ [4] by tuning the probe photon energy to the polaronic absorption peak. Although we used polaronic signatures in both measurements, associated fast relaxation processes in transient reflectivity signal have never been completely understood despite being essential for description of dynamic properties of Jahn-Teller polarons that exist in manganites in both paramagnetic and ferromagnetic phases. To obtain further information on the sub-picosecond polaron dynamics, we performed ultrafast optical reflectivity measurements as a function of temperature ($T$) in the systems with rather simple phase diagrams, such as three-dimensional manganite $La_{0.7}Ca_{0.3}MnO_3$ (LCMO) and the quasi-two dimensional bilayered manganite $La_{1.4}Sr_{1.6}Mn_2O_7$ (bilayered LSMO). The observed recovery process in the high temperature paramagnetic insulating phase corresponds to the redressing of the photoexcited electron to its polaronic ground state. A comparison of the temperature dependence of the relaxation times in LCMO with those in bilayered LSMO suggests that the recovery time of the polaron is strongly influenced by the dimensionality of the electronic structure and spin orientation [5]. Furthermore, we observe a crossover to a more conventional electron-phonon relaxation in the ferromagnetic metallic phases.

LCMO, which has a perovskite structure, is a ferromagnetic metal below $T_C$ =250 K and a paramagnetic insulator above $T_C$. Similarly, bilayered LSMO has a 3D ferromagnetic transition near 90 K and features a layered perovskite structure which consists of quasi-2D $MnO_2$ bilayers separated by insulating blocks of $(La, Sr)_2 O_2$ [14] . The samples we have investigated are single crystals with surface oriented normal to *c* direction. We have employed a standard optical pump-probe setup where Ti:Sapphire laser produces 20-fs pulses with 80 MHz repetition rate that are centered at 1.5 eV photon energy. The photoinduced change in reflectivity *ΔR/R* was measured using an avalanche photodiode and lock-in detection. The pump fluence was kept below 10 $\mu J/cm^2$ to minimize heating but still produce peak values of *ΔR/R* of approximately $4 \times 10^{-5}$. Figures 2(a) and (b) present *ΔR/R* as a function of time at various temperatures for LCMO and bilayered LSMO, respectively. Small oscillation around 1*ps* (clearly seen in case of LCMO 450K) could be due to Raman active coherent phonon oscillation [15] and this indicates that the lattice system does not reach thermal equilibrium of electron-lattice coupled system via the mutual thermal contact [15]. The following slow decay part would be corresponding to the cooling process. The solid lines in figure 2 represent best fits to a two exponential decay model, which includes a longer recovery associated with spin-lattice thermalization..

Our measurements of this longer spin-lattice thermalization on the single crystal LCMO agree with previous data acquired on thin film samples [11] and, in what follows, we will not further discuss this component of the dynamics in LCMO and bilayered LSMO. Interestingly, both materials exhibit a sub-ps recovery with a peak transient reflectivity amplitude being largest in the vicinity of $T_C$ and becoming smaller in both the ferromagnetic and paramagnetic phases (Fig. 3(a) shows this for LSMO). We note that similar behavior is also apparent in the temperature dependence of the neutron and x-ray scattering experiments [7, 8, 9] as shown by overlaying neutron data with our optical *ΔR/R* results in Fig. 3(a). Such a correlation between *ΔR/R* amplitude and neutron scattering experiment has been observed before in other manganites using mid-IR probe resonant with polaron energy [4]. Similarly, the fast relaxation time constant τ of both compounds display a pronounced temperature dependence with the peak at $T_C$ that will be discussed in detail below.

As we described in Eq.1, below $T_C$ the effective hopping matrix element $t_{ij}$ becomes dominant because ferromagnetically aligned local spin maximizes $t_{ij}=t\ cos(\theta_{ij}/2) =t$ ($\theta_{ij}=0$ degree) and overcomes electron localization due to the lattice distortion described by the second term in Eq. 1. Therefore, initial fast relaxation at $T < 0.9T_C$ should correspond to the time for mobile quasiparticle to shed its excess energy through electron-phonon scattering [4, 11, 16]. 17

The low temperature behavior is reminiscent of regular metals such as Au and Ag [17]. In particular, the excited nonequilibrium electron distribution initially thermalizes via electron-electron scattering and then relaxes through the emission of incoherent acoustic and optical phonons. The independence of the relaxation time at the temperatures below approximately $0.8T_C$ is consistent with electron relaxation of typical metal as shown in ref. [17]. []. However, near $T_C$ the relaxation time in both cases shows a sudden jump in the magnitude of τ. Above $T_C$ localized polaronic term in Eq.1 gradually overcomes the hopping amplitude and causes τ to increase monotonically with decreasing temperature - in other words, upon approaching $T_C$ from above, the time for the photoexcited electron to redress to a polaronic state increases. At room temperature (300 K, $T/T_C \sim$ 3.3 for bilayered LSMO and $T/T_C \sim$ 1.2 for LCMO), τ is ~ 0.23 ps (0.35 ps) for LCMO (bilayered LSMO). This coincides with the increasing importance of polaronic effects in these materials near and above $T_C$. Therefore, we argue that the relaxation dynamics we measure for $T > 0.9T_C$ (in both LCMO and bilayered LSMO) are dominated by polaronic effects where, in particular, the measured relaxation dynamics now correspond to the time for a photoexcited electron to redress to a polaronic state. This is in contrast to the case for $T < 0.9T_C$, which, as just discussed, corresponds to the time for a free quasiparticle to shed its excess energy through e-ph relaxation.

The optical conductivity of LCMO and bilayered LSMO shows multiple broad absorption peaks in the near-infrared region around 1.2 eV [18,19]. The $Mn^{3+}$ $e_g$ levels with four $d$ electrons are JT-split into $e_{1g}$ and $e_{2g}$, while the $Mn^{4+}$ levels are not (Fig. 1). Therefore, near-infrared band consists of the sum of two bands [20]; one band with a peak below 1 eV which is due to intersite electron hopping from $e_{1g}$ ($Mn^{3+}$) to $e_g$ ($Mn^{4+}$) and the other band with a peak around 1.5 eV which is due to either on-site or inter-site $Mn^{3+}$ $d-d$ transitions [20-22]( Jung *et al*, [20] pointed out that intersite transion may move to a higher energy which is very likely to move beyond the energy range we interested ). When the sample is excited by the pump pulse at 800 nm (∼1.5 eV), a polaronic carrier absorbs a photon and is excited to the upper $Mn^{3+}$ $e_{2g}$ level. An electron in this upper $Mn^{3+}$ $e_{2g}$ level is unstable and will relax to a lower level. There are two pathways for this relaxation; either directly back to the $Mn^{3+}$ $e_{1g}$ state or through the nearest-neighbor $Mn^{4+}$ $e_g$ level. The direct on-site relaxation to $Mn^{3+}$ $e_{1g}$ can occur on a very fast timescale without any need for motion of the oxygen atoms. There can also be a transition to a $Mn^{4+}$ $e_g$ level, which will first occur without lattice deformation. However, the occupation of an initially degenerate $e_g$ level of $Mn^{4+}$ will result in lattice motion and the electron will be localized to remove the degeneracy of the $Mn^{3+}$ $e_g$ level through a JT distortion of the $MnO_6$ octahedron.

Emin and Holstein [23] showed that a scaling analysis of the adiabatic eigenstates of an electron placed in a deformable continuum enables us to obtain information about the polaronic bound state in a rather simple way. Following their paper, for the system with short range electron-phonon interactions, which describes best the interaction of electron with localized JT distortion in manganites, its energy could be expresses as [23]

$$E(R) = \frac{T_e}{R^2} - \frac{V^S_{int}}{2R^d},  \qquad (2)$$

where $T_e$, $V^S_{int}$ and $d$ are the electron's kinetic energy, the constant for the short range interactions and the dimensionality of the system, respectively. $R$ is the scaling factor for the electron wave function $R^{-d/2}\Psi(\vec{r}/R)$, chosen so that the energy (as a function of $R$) of a finite radius eigenstate must have its minimum at the $R=1$, i.e. the scale corresponding to the actual eigenstate $\Psi(\vec{r})$. One important thing to keep in mind is that the energy expression in Eq. 2 is cut off at a small $R=R_c$ due to atomistic limits, with $E(R_c)=E_{sp}$ being an energy of small polaron as described in Ref. 23. Importantly, small polaron minimum in Eq.2 is stable only when its energy is smaller than free electron minimum $E(R = \infty) = 0$, i.e. when $E_{sp} < 0$.

For $T<T_C$, the effective hopping matrix element $t_{ij}$ becomes maximum because of ferromagnetically aligned neighboring spins. If $E(R_c) >0$, the small polaron state is not stable and the system has a free electron ground state as shown for $T_e = 1$ in Figure 4(a). In contrast, for $T > T_C$, $t_{ij}$ becomes smaller due to thermally disordered spins, which effectively reduce hopping probability. When temperature increases (corresponding to lower $T_e$ in Eq. 2), the energy at the $R_c$ becomes negative and it is now possible to have the polaronic ground state as shown for $T_e = 0.7$ in Figure 4(a) red line. When we compare the relaxation of excited states after applying an external perturbation, such as the pump laser pulse, at $T<T_C$ and $T>T_C$, the recovery time to the original state (before perturbation) is longer at $T>T_C$ because there is a barrier that has to be overcome in order to return to the polaronic ground state (Figure 4(b)).

When temperature is increased, ground state gradually shifts from a free electron to a small polaron with a negative $E(R_c)$, where an energy barrier begins to affect the relaxation and leads to a sharp increase of redressing time at $T_C$ as shown in Figure 3(b) for LCMO. The relaxation time of the electron-lattice system in the paramagnetic insulating phase corresponds to the recovery time of an initially excited electron back to polaronic state. When we further increase the temperature in the $T>T_C$ region, the electron kinetic energy $T_e$ becomes even smaller. Reduced electron kinetic energy decreases the height of an energy barrier (based on Eq. 2 and by comparison the red line with the blue line in figure 4(a)) and also reduces the energy at $R_c$. This results in the shorter redressing time necessary for returning to the polaronic bound state. Thus, the negative slope in our data at $T>T_C$ in Figure 3(b) could be explained with this qualitative Emin's analysis [23].

It is interesting to compare our results with the resistivity data for different dimensions of manganese oxides in figure 2(a) in Moritomo *et al*. (ref. [2]). Their experimental data shows that steep rise of resistivity in 2 dimensional manganese oxides are observed around $T_C$ comparing with rather gradual resistivity change in 3 D. Since both electron-phonon interaction and kinetic energy of electron ($T_e$) influence resistivity and overall values of resistivity are larger in 2D, the data indicate that 2D would have more electron-phonon coupling than 3D. However, origins of strong electron phonon coupling in reduced dimensionality are not clear as follows

In the case of 2D manganite like LSMO, Eq. 2 with $d = 2$ shows that there is no bound polaronic state). However, experimental data show the presence of 2D polaronic state in this material [24]. It is expected that the polaron size in the layered perovskite is larger than in the 3D perovskite due to the reduced dimensionality [5, 25]. Ku et al. reported that the correlation function decays more rapidly and that the surrounding phonons are more localized near the electron in higher dimensions [5]. Heffner et al. also observed larger magnetoelastic polarons in the 2D layered system than in the 3D perovskites [25]. Equation (2) was written with an assumption that the phonon mass equals to zero. However, including the finite phonon mass and applying more exact calculation using a variation method show that 2D polaronic bound state indeed exists [5,26] (actually, it appears in all dimensions including 4D). These calculations also predict that 2D polaronic state would have lower electron kinetic energy and smaller $V^S_{int}$ magnitude than 3D state, and result in smaller energy barrier between polaronic and free electron ground states. Following these arguments, 2D polaronic redressing time should be effectively shorter than the 3D, which contradicts our observations shown in Figure 3 (b).

In order to explain this controversy, we must also consider the spin ordering as there is a delicate interplay between the lattice and spin degrees of freedom in the manganites. For bilayered LSMO, magnetization measurements above $T_C$ indicate an emergence of in-plane, short range 2D ferromagnetic correlations, since the magnetization shows a plateau between 90 and 250 K [27]. A pair density function analysis of neutron scattering data by Louca et al.[28] also indicates the occurrence of double exchange processes at temperatures much higher than the 3D ferromagnetic ordering temperature ($T_c$ = 90 K) of the layered system [28]. Thus, while complete 3D spin ordering occurs below $T_c$, there is considerable in-plane spin ordering that manifests well above $T_C$. This affects the measured polaron redressing time in the following manner: As the temperature is decreased, the strengthening of 2D spin correlations facilitates increased hopping of the photoexcited carriers through double exchange processes. An increased hopping rate corresponds to a decrease in the average time an electron spends at a given Mn site. If this reduces the probability of JT octahedral distortion which effectively increases the time it takes to reform the polaronic state, 2D redressing time should be larger than 3D.

To provide consistent explanation, we have to consider yet another effect of lower dimensionality in bilayered LSMO. Jahn-Teller distortion with polaron formation in lower dimensions was questioned before due to additional energy splitting caused by spatial confinement in 2D layers [2]. In our previous discussion of the bilayer LSMO compound, we completely neglected the fact that $e_g$ states may have already been split (beside JT distortion) due to lower crystal field symmetry in the layered compound. This makes a direct comparison with a 3D cubic LCMO compound less obvious. However, if energy splitting of $e_g$ states solely by the low dimensionality is stronger than JT-induced splitting with ensuing energy barrier between the polaronic and free electron ground states being bigger, the redressing process in 2D would take longer time than in 3D manganites.

However, it is beyond the scope of this paper to examine all the effects of lower dimensionality, including quantum confinement and changes in crystal fields, on the relaxation processes in manganites. Experimentally ultrafast spectroscopy is sensitive to measure polaron redressing time as shown in the paper and should be carried the same kind of experiments in thin film manganese oxides to understand inter-correlation between JT distortion and $e_g$ split with lower dimensionality.

In conclusion, we have compared and contrasted the quasiparticle dynamics in 3D LCMO and quasi-2D bilayered LSMO single crystals in a broad temperature range. We observed the release of photoexcited electron from the polaronic confinement with subsequent redressing of electron to the polaronic or free electron ground state, which occurs on sub-picosecond timescales in in both materials. A comparison of the temperature dependence of the relaxation time between LCMO and bilayered LSMO in the paramagnetic insulating phase at $T > T_c$ suggests that simple polaronic redressing picture have to be significantly modified in the 2D materials and a new model should be introduced where polaron redressing time is strongly influenced by the dimensionality of the electronic structure and the presence of spin correlations above $T_C$.

We thank M. J. Graf, K. H. Ahn, Y. K. Bang, and K. H. Kim for helpful discussions. The work at Los Alamos was supported by the DOE Laboratory Directed Research and Development Program.

REFERENCES

[1] A. J. Millis, P. B. Littlewood, and B. I. Shraiman, Phys. Rev. Lett. 74, 5144 (1995); H. Röder, J. Zang, and A. R. Bishop, Phys. Rev. Lett. 76, 1356 (1996).

[2] S. Jin, T. H. Tiefel, M. McCormack, R. A. Fastnacht, R. Ramesh, and L. H. Chen, Science 264, 413 (1994); Y. Moritomo, Y. Tomioka, A. Asamitsu, Y. Tokura and Y. Matsui, Nature 380 141 (1996); M. B. Salamon and M. Jaime, Rev. Mod. Phys. 73, 583 (2001).

[3] C. Zener, Phys. Rev. 82, 403 (1951); J. B. Goodenough, Phys. Rev. 100, 564 (1955); P. W. Anderson and H. Hasegawa, Phys. Rev. 100, 675 (1955).

[4] R. P. Prasankumar, S. Zvyagin, K. V. Kamenev, G. Balakrishnan, D. Mck. Paul, A. J. Taylor, and R. D. Averitt, Phys. Rev (R). 76 20402 (2007)

[5] Li-Chung Ku, S. A. Trugman, and J. Bonca, Phys. Rev. B 65, 174306 (2002).

[6] D. Khomskii, Electronic structure, exchange, and magnetism in oxides. In *Lecture Notes on Physics*, ed. Thrnton, M. J. & Ziese, M. (Springer-Verlag, Berlin, 2001)

[7] S. Shimomura, N. Wakabayash, H. Kuwahara and Y. Tokura, Phys. Rev. Lett. 83, 4389 (1999)

[8] L. Vasiliu-Doloc, S. Rosenkranz, R. Osborn, S. K. Sinha, J. W. Lynn, J. Mesot, O. H. Seeck, G. Presoti, A. J. Fedro and J. F. Mitchell, Phys. Rev. Lett. 83, 4393 (1999)

[9] D. N. Argyriou, J. W. Lynn, R. Osborn, B. Campbell, J. F. Mitchell, U. Ruett, H. N. Bordallo, A. Wildes, and C. D. Ling, Phys. Rev. Lett. 89, 36401 (2002)

[10] K. Matsuda, A. Machida, Y. Moritomo, and A. Nakamura, Phys. Rev. B 58 R4203 (1998).

[11] R. D. Averitt, A. I. Lobad, C. Kwon, S. A. Trugman, V. K. Thorsmolle, and A. J.

Taylor, Phys. Rev. Lett. 87, 017401 (2001).

[12] T. Ogasawara, T. Kimura, T. Ishikawa, M. Kuwata-Gonokami, and Y. Tokura, Phys. Rev. B 63, 113105 (2001).

[13] J. Lee, S.A. Truggman, C. D. Batista, C. L. Zhang, D. Talbayev, X. S. Xu, S. –W. Cheong, D. A. Yarotski, A. J. Taylor, and R. P. Prasankumar, Sci. Rep. **3**, 2654 (2013)

[14] T. G. Perring, G. Aeppli, Y. Moritomo, and Y. Tokura, Phys. Rev. Lett. 78, 3197 (1997)

[15] Takeshi Ogasawara, Katsuhiro Tobe, Tsuyoshi Kimura, Hiroshi Okamoto, and Yoshinori Tokura, J. Phys. Soc. Jpn. 71, 2380 (2002)

[16] Ahmed I. Lobad, Richard D. Averitt, and Antoinette J. Taylor, Phys. Rev. B. 63, 060410(R) (2000)

[17] R. H. M. Groeneveld, R. Sprik, and A. Lagendijk, Phys. Rev. B 51, 11433 (1995); G. L. Eesley Phys. Rev. Lett. 51, 2140 (1983).

[18] K. H. Kim, J. H. Jung, and T.W. Noh, Phys. Rev. Lett. 81, 1517 (1998).

[19] H. J. Lee, K. H. Kim, J. H. Jung, T. W. Noh, R. Suryanarayanan, G. Dhalenne, and A. Revcolevschi, Phys. Rev. B 62, 11320 (2000); T. Ishikawa, K. Tobe, T. Kimura, T. Katsufuji, Y. Tokura, Phys. Rev. B 62, 12354 (2000);

[20] J. H. Jung, K. H. Kim, T. W. Noh, E. J. Choi, and J. Yu, Phys. Rev. B 57, R11043 (1998); M. Quijada, J. Černe, J. R. Simpson, H. D. Drew, K. H. Ahn, A. J. Millis, R. Shreekala, R. Ramesh, M. Rajeswari, and T. Venkatesan, Phys. Rev. B 58, 16093 (1998); A. Machida, Y. Moritomo, and A. Nakamura, Phys. Rev. B 58, 12540 (1998).

[21] P. B. Allen and V. Perebeinos, Phys. Rev. Lett. 83, 4828 (1999).

[22] There is an argument about the exact origin of the $e_{1g}$($Mn^{3+}$) to $e_{2g}$($Mn^{3+}$) transition around 1.5 eV. Jung *et al.* [20] assigned it to an intra-atomic transition between these JT split $Mn^{3+}$ levels

while Quijada *et al.* [20] assigned it to an interatomic transition between the same levels. On the other hand, Machida et al. [20] attributed it to an optical transition related to JT clusters. Allen *et al.* [13] showed that 1.5 eV peak can be explained in terms of a Franck-Condon series of the intra-atomic $e_{1g}$(Mn$^{3+}$) to $e_{2g}$(Mn$^{3+}$) transition.

[23] D. Emin, and T. Holstein, Phys. Rev. Lett. **36**, 323 (1976)

[24] D. N. Argyriou, J. W. Lynn, R. Osborn, B. Campbell, J. F. Mitchell, U. Ruett, H. N. Bordallo, A. Wildes, and C. D. Ling Phys. Rev. Lett. **89**, 036401(2002)

[25] R. H. Heffner, D. E. MacLaughlin, G. J. Nieuwenhuys, T. Kimura, G. M. Luke, Y. Tokura, and Y. J. Uemura, Phys. Rev. Lett. 81, 1706 (1998).

[26] D. Emin, Polarons, Cambridge Univ Press, 2013

[27] T. Kimura, A. Asamitsu, Y. Tomioka, and Y. Tokura, Phys. Rev. Lett. 79, 3720 (1997).

[28] D. Louca, G. H. Kwei, and J.F. Mitchell, Phys. Rev. Lett. 80, 3811 (1998).

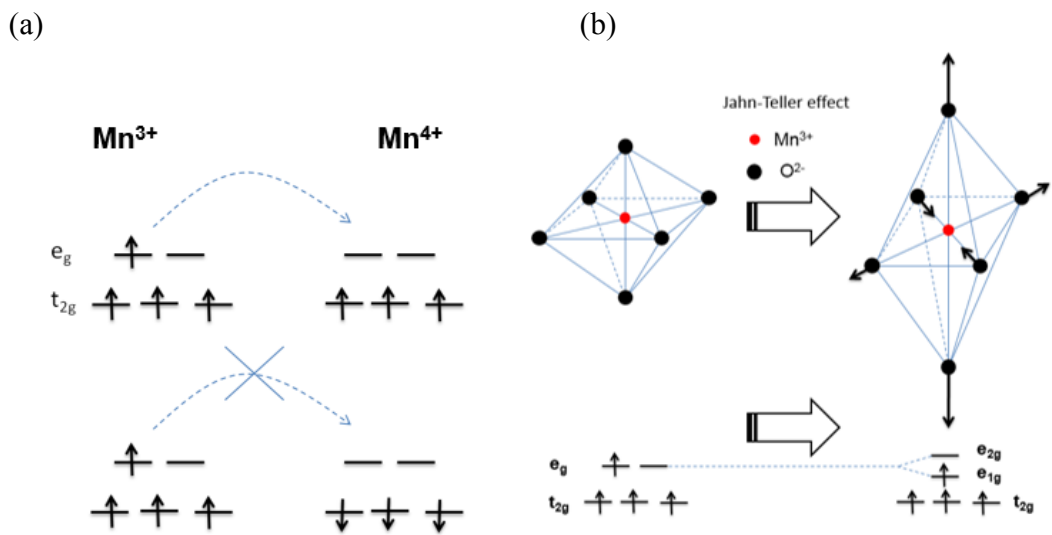

**Figure 1.** (a) Schematic of the double exchange interaction: If $Mn^{3+}$ is aligned ferromagnetically with $Mn^{4+}$, a charge transfer from $Mn^{3+}$ to $Mn^{4+}$ is allowed. However if these two ions are aligned antiferromagnetically, a charge transfer between $Mn^{3+}$ to $Mn^{4+}$ gives a high energy state which is energetically forbidden by Hund's rule of maximum on-site spin. (b) The $Mn^{3+}O_6$ octahedron distortion associated with the Jahn-Teller effect and schematic energy level alignment of the Mn ion. It shows that Jahn-Teller distortion splits two-fold degenerate $e_g$ levels.

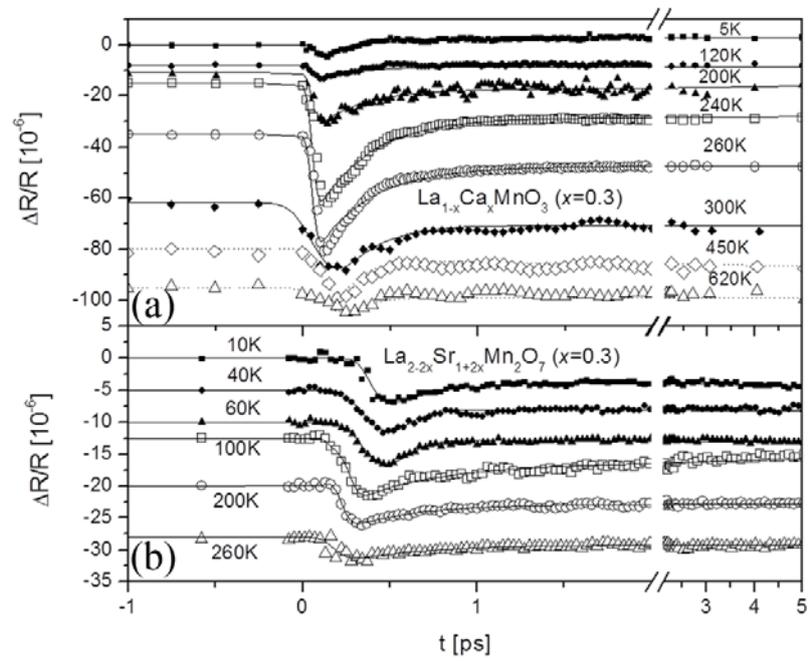

**Figure 2.** Temperature-dependent relaxation dynamics of (a) LCMO and (b) bilayered LSMO.

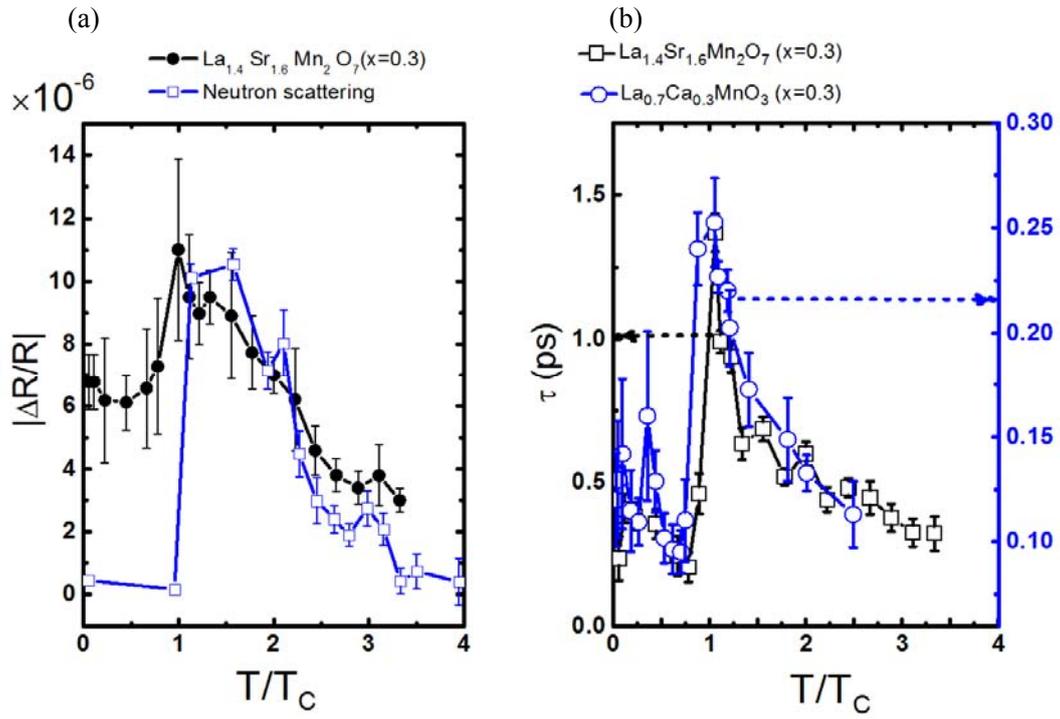

Figure 3 (a) Amplitude of Δ$R$/$R$ measured at the peak as compared to the elastic neutron scattering amplitude [24], and (b) temperature dependent relaxation time of LSMO (black) and LCMO (blue).

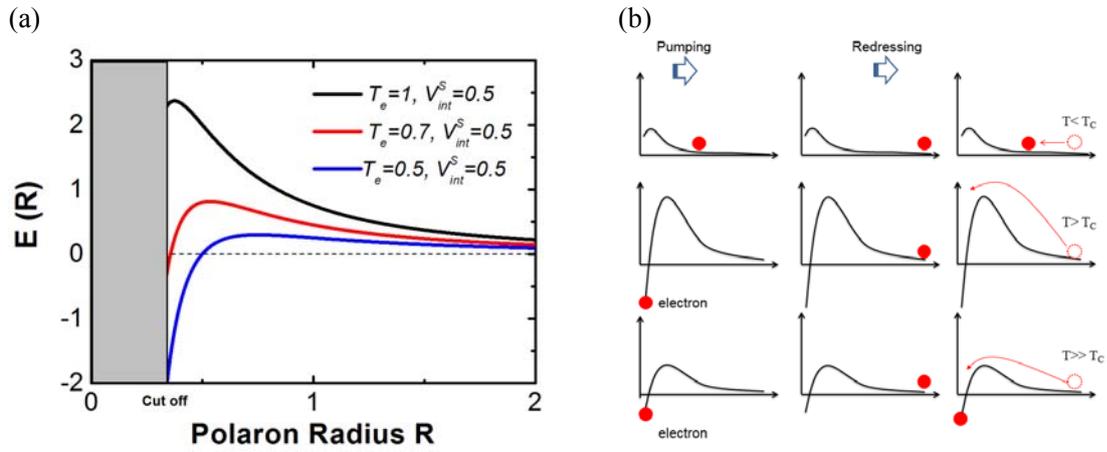

Figure 4. (a) Representation of Eq. 2 for two $T_e$ values. Curves stop at cut off radius shown as black area. (b) Schematic pathway of quasiparticle relaxation following photoexcitation for $T<T_C$, $T>T_C$ and $T>>T_C$